\documentclass[12pt]{article}
\usepackage{amsmath,amssymb,amsthm,amsxtra,overpic,bbm,bm,epsfig,ulem}
\usepackage{color}
\usepackage{cite}

\usepackage{hyperref}
\usepackage{url}

\def\thefootnote{\fnsymbol{footnote}}
\allowdisplaybreaks[4]

\begin{document}

\vspace{0.2cm}

\begin{center}
{\large\bf On the orthogonal matrix in the Casas-Ibarra parametrization for \\
the Yukawa interactions of Majorana neutrinos}
\end{center}

\vspace{0.2cm}

\begin{center}
{\bf Zhi-zhong Xing$^{1,2}$}
\footnote{E-mail: xingzz@ihep.ac.cn}
\\
{\small $^{1}$Institute of High Energy Physics and School of Physical Sciences, \\
University of Chinese Academy of Sciences, Beijing 100049, China \\
$^{2}$Center of High Energy Physics, Peking University, Beijing 100871, China}
\end{center}

\vspace{2cm}
\begin{abstract}
The Casas-Ibarra (CI) parametrization of the Yukawa coupling matrix of Majorana
neutrinos is generalized by considering the exact seesaw relation and including
non-unitarity of the $3 \times 3$ Pontecorvo-Maki-Nakagawa-Sakata (PMNS) flavor
mixing matrix. With the help of a full $6 \times 6$ Euler-like block description
of the flavor structure for the seesaw mechanism, we find that the orthogonal
matrix $\mathbb{O}$ in the CI parametrization can be expressed as $\mathbb{O}^{}_{ij} =
\sqrt{M^{}_j/m^{}_i} \hspace{0.05cm} F^{}_{ij}$ with $m^{}_i$ and $M^{}_j$ being
the masses of light and heavy Majorana neutrinos and $F^{}_{ij}$ consisting of
the PMNS and active-sterile flavor mixing parameters (for $i, j = 1, 2, 3$).
Assuming a specific pattern of $\mathbb{O}$ is therefore equivalent to imposing some
special conditions on the seesaw parameter space.
\end{abstract}

\newpage

\def\thefootnote{\arabic{footnote}}
\setcounter{footnote}{0}

\setcounter{equation}{0}
\section{Introduction}

The canonical seesaw mechanism has widely been accepted as a most natural and
most economical extension of the standard model (SM) of particle physics for
the purpose of understanding why the masses of the three known neutrinos
($\nu^{}_1$, $\nu^{}_2$ and $\nu^{}_3$ versus $\nu^{}_e$, $\nu^{}_\mu$ and
$\nu^{}_\tau$) are so small as compared with the masses of their charged
counterparts (i.e., $e$, $\mu$ and $\tau$)~\cite{Minkowski:1977sc,
Yanagida:1979as,GellMann:1980vs,Glashow:1979nm,Mohapatra:1979ia}. In this
mechanism the flavor sector of the SM is modified by introducing three
right-handed neutrino fields $N^{}_{\alpha \rm R}$ (for $\alpha = e, \mu, \tau$)
and their kinetic terms, adding a lepton-number-violating Majorana mass term
constituted with $N^{}_{\alpha \rm R}$ and their charge-conjugated fields
$\left(N^{}_{\alpha \rm R}\right)^{\rm c}$, and allowing for the Yukawa
interactions between the SM Higgs doublet $H$ and the chiral neutrino
fields~\cite{Xing:2009in}. As masses of the three heavy neutrinos (denoted
as $M^{}_i$ for $N^{}_i$ with $i = 1, 2, 3$) have nothing to do with the
electroweak symmetry breaking scale of the SM that is characterized by the
Higgs vacuum expectation value $\langle H\rangle \simeq 174~{\rm GeV}$, their
values can be far above $\langle H\rangle$. Integrating out the heavy degrees
of freedom will therefore give rise to small masses of
${\cal O}(\langle H\rangle^2/M^{}_i)$ for the active neutrinos $\nu^{}_i$ (for
$i = 1, 2, 3$), in agreement with the spirit of Weinberg's SM effective field
theory~\cite{Weinberg:1967tq}. Moreover, the CP-violating and out-of-equilibrium
decays of three heavy Majorana neutrinos $N^{}_i$ (for $i = 1, 2, 3$) result in
a net lepton-antilepton asymmetry in the early Universe~\cite{Fukugita:1986hr}
--- the so-called {\it leptogenesis} mechanism which may lead to
{\it baryogenesis} as a viable explanation of the cosmic baryon-antibaryon
asymmetry~\cite{ParticleDataGroup:2024cfk}.

It is the SM-like Yukawa interactions that dictate the decays of heavy Majorana
neutrinos and bridge the gap between the active (light) and sterile (heavy)
sectors of the seesaw framework, but their flavor structure is completely
undetermined and thus obstructs our specific studies of the seesaw-plus-leptogenesis
phenomenology. For this reason, Casas and Ibarra have proposed a working
parametrization of the Yukawa coupling matrix $Y^{}_\nu$ in terms of the diagonal
neutrino mass matrices $D^{}_\nu \equiv {\rm Diag}\{m^{}_1, m^{}_2, m^{}_3\}$ of
$\nu^{}_i$ and $D^{}_N \equiv {\rm Diag}\{M^{}_1, M^{}_2, M^{}_3\}$ of $N^{}_i$
together with the $3 \times 3$ unitary Pontecorvo-Maki-Nakagawa-Sakata (PMNS)
lepton flavor mixing matrix $U^{}_0$~\cite{Pontecorvo:1957cp,Maki:1962mu,
Pontecorvo:1967fh} and an arbitrary complex orthogonal matrix $\mathbb{O}$ in the
{\it diagonal} charged-lepton flavor basis~\cite{Casas:2001sr}
\footnote{Note that the notations used here are different from those adopted in
Ref.~\cite{Casas:2001sr}, as our Lagrangian for the seesaw mechanism in Eq.~(\ref{2})
is expressed in a different convention.}:
\begin{eqnarray}
Y^{}_\nu \simeq \frac{\rm i}{\langle H\rangle} U^{}_0 \sqrt{D^{}_\nu} \hspace{0.05cm}
\mathbb{O} \sqrt{D^{}_N} \; ,
\label{1}
\end{eqnarray}
which is consistent with the popular but approximate seesaw
relation~\cite{Minkowski:1977sc,Yanagida:1979as,GellMann:1980vs,Glashow:1979nm,
Mohapatra:1979ia}. Such a parametrization resembles the QCD factorization in hadron
physics~\cite{Collins:1989gx,Beneke:1999br,Beneke:2000ry} to some extent, where the
``perturbation" parts containing $U^{}_0$, $D^{}_\nu$ and $D^{}_N$ have been factored
out and the unknown ``non-perturbation" part is formally described by $\mathbb{O}$.
In this case the flavor structure of $Y^{}_\nu$ can largely be fixed if a specific
pattern of $\mathbb{O}$ is assumed, as in many phenomenological applications (see, e.g.,
Refs.~(\cite{Ibarra:2003up,Pascoli:2006ie,Blanchet:2006be,Molinaro:2009lud,Xing:2012gd,
Bambhaniya:2016rbb,Xing:2020erm,Xing:2020ghj,Zhao:2021dwc,Granelli:2023tcj,Zhao:2024uid,
Zhao:2024mrs,Shao:2024fem}).

But the physical meaning of the orthogonal matrix $\mathbb{O}$ in Eq.~(\ref{1}) is not
so clear cut. In particular, a straightforward observation that $Y^\dagger_\nu Y^{}_\nu$
is apparently independent of $U^{}_0$ and hence the {\it unflavored} leptogenesis
is not correlated with leptonic CP violation at low energies has been
questioned~\cite{Xing:2009vb}, and whether possible non-unitarity of the $3 \times 3$
PMNS matrix may offer a way out of the above observation has been
discussed~\cite{Rodejohann:2009cq,Antusch:2009gn}. What does $\mathbb{O}$ really mean
in the Casas-Ibarra (CI) parametrization?

The present work aims to answer this question. First of all, we are going to
generalize the CI parametrization by starting from the exact seesaw
relation and taking into account possible non-unitarity of the PMNS matrix.
Secondly, we shall derive the complex orthogonal matrix $\mathbb{O}$ in this
parametrization with the help of a full $6 \times 6$ Euler-like block description
of the flavor structure for the canonical seesaw mechanism. We find
$\mathbb{O}^{}_{ij} = \sqrt{M^{}_j/m^{}_i} \hspace{0.05cm} F^{}_{ij}$ with $m^{}_i$
and $M^{}_j$ being the masses of light and heavy Majorana neutrinos and $F^{}_{ij}$
consisting of the PMNS and active-sterile flavor mixing parameters (for
$i, j = 1, 2, 3$). So assuming a specific pattern of $\mathbb{O}$ is actually equivalent
to imposing some special conditions on the seesaw parameter space.

\section{A generalized CI parametrization}

Without loss of any generality, let us write out the leptonic Yukawa interactions
and the Majorana neutrino mass term of the canonical seesaw mechanism in the
basis where both the charged-lepton Yukawa coupling matrix
and the right-handed neutrino mass matrix are diagonal and real:
\begin{eqnarray}
-{\cal L}^{}_{\rm SEESAW} = \overline{\ell^{}_{\rm L}} \hspace{0.05cm} Y^{}_l
H \hspace{0.02cm} l^{}_{\rm R} + \overline{\ell^{}_{\rm L}} \hspace{0.05cm}
Y^{}_\nu \widetilde{H} N^{}_{\rm R}
+ \frac{1}{2} \hspace{0.05cm}
\overline{(N^{}_{\rm R})^{\rm c}} \hspace{0.05cm} D^{}_{\rm R} N^{}_{\rm R}
+ {\rm h.c.} \; ,
\label{2}
\end{eqnarray}
in which $\ell^{}_{\rm L}$ denotes the leptonic $\rm SU(2)^{}_{\rm L}$ doublet of
the SM, $\widetilde{H} \equiv {\rm i} \sigma^{}_2 H^*$ with $H$ being the SM Higgs
doublet and $\sigma^{}_2$ being the second Pauli matrix, $l^{}_{\rm R}$
and $N^{}_{\rm R}$ stand respectively for the column vectors of the right-handed
charged lepton and neutrino fields which are the $\rm SU(2)^{}_{\rm L}$
singlets, $(N^{}_{\rm R})^{\rm c} \equiv {\cal C} \overline{N^{}_{\rm R}}^T$
with $\cal C$ being the charge-conjugation matrix, $Y^{}_l$ denotes the diagonal
and real charged-lepton Yukawa coupling matrix, $Y^{}_\nu$ represents the arbitrary
Yukawa coupling matrix of massive neutrinos, and $D^{}_{\rm R}$ is the diagonal
and real right-handed neutrino mass matrix. After spontaneous electroweak gauge
symmetry breaking, Eq.~(\ref{2}) becomes
\begin{eqnarray}
-{\cal L}^{}_{\rm m} = \overline{l^{}_{\rm L}} \hspace{0.05cm}
D^{}_l \hspace{0.03cm} l^{}_{\rm R} + \frac{1}{2} \hspace{0.05cm}
\overline{\big[\begin{matrix} \nu^{}_{\rm L} & \hspace{-0.1cm}
(N^{}_{\rm R})^{\rm c}\end{matrix}
\big]} \left(\begin{matrix} {\bf 0} &  M^{}_{\rm D} \cr
M^T_{\rm D} & D^{}_{\rm R} \end{matrix}\right)
\left[\begin{matrix} (\nu^{}_{\rm L})^{\rm c} \cr N^{}_{\rm R} \end{matrix}\right]
+ {\rm h.c.} \; ,
\label{3}
\end{eqnarray}
where $l^{}_{\rm L}$ and $\nu^{}_{\rm L}$ stand respectively for the column
vectors of the left-handed charged-lepton and neutrino fields,
$D^{}_l \equiv Y^{}_l \langle H\rangle$ and
$M^{}_{\rm D} \equiv Y^{}_\nu \langle H\rangle$ are defined.

To diagonalize the symmetric $6\times 6$ neutrino mass matrix in Eq.~(\ref{3}),
we make the Autonne-Takagi transformation~\cite{Autonne1915,Takagi1924} as follows:
\begin{eqnarray}
\mathbb{U}^\dagger \left ( \begin{matrix} {\bf 0} & M^{}_{\rm D}
\cr M^T_{\rm D} & D^{}_{\rm R} \cr \end{matrix} \right )
\mathbb{U}^* = \left( \begin{matrix} D^{}_\nu & {\bf 0} \cr {\bf 0} &
D^{}_N \cr \end{matrix} \right) \; ,
\label{4}
\end{eqnarray}
where $\mathbb{U}$ is a $6\times 6$ unitary matrix,
$D^{}_\nu \equiv {\rm Diag}\big\{m^{}_1, m^{}_2, m^{}_3\big\}$ and
$D^{}_N \equiv {\rm Diag}\big\{M^{}_1, M^{}_2, M^{}_3 \big\}$ have been
defined for the masses of light and heavy Majorana neutrinos. It is
instructive to decompose $\mathbb{U}$ into the product of three matrices,
\begin{eqnarray}
\mathbb{U} = \left( \begin{matrix}
I & {\bf 0} \cr {\bf 0} & U^{\prime}_0 \cr \end{matrix} \right)
\left( \begin{matrix} A & R \cr S & B \cr \end{matrix} \right)
\left( \begin{matrix} U^{}_0 & {\bf 0} \cr {\bf 0} & I \cr
\end{matrix} \right) \; ,
\label{5}
\end{eqnarray}
where the $3\times 3$ unitary matrix $U^\prime_0$ is primarily related to the sterile
(heavy) neutrino sector, $U^{}_0$ denotes the other $3\times 3$ unitary matrix that
is primarily responsible for flavor mixing in the active (light) neutrino sector,
and the $3\times 3$ matrices $A$, $B$, $R$ and $S$ signify the interplay between these two
sectors~\cite{Xing:2007zj,Xing:2011ur,Xing:2020ijf}. The unitarity of $\mathbb{U}$
assures
\begin{eqnarray}
&& A A^\dagger + R R^\dagger = B B^\dagger + S S^\dagger = I \; ,
\nonumber \\
&& A S^\dagger + R B^\dagger = A^\dagger R + S^\dagger B = {\bf 0} \; ,
\nonumber \\
&& A^\dagger A + S^\dagger S = B^\dagger B + R^\dagger R = I \; ;
\hspace{1.4cm}
\label{6}
\end{eqnarray}
and the transformation made in Eq.~(\ref{4}) leads us to the weak charged-current
interactions of light and heavy Majorana neutrinos in the mass basis:
\begin{eqnarray}
-{\cal L}^{}_{\rm cc} = \frac{g}{\sqrt{2}} \hspace{0.1cm}
\overline{\big(\begin{matrix} e & \mu & \tau\end{matrix}\big)^{}_{\rm L}}
\hspace{0.1cm} \gamma^\mu \left[ U \left( \begin{matrix} \nu^{}_{1}
\cr \nu^{}_{2} \cr \nu^{}_{3} \cr\end{matrix} \right)^{}_{\hspace{-0.08cm} \rm L}
+ R \left(\begin{matrix} N^{}_{1} \cr N^{}_{2} \cr N^{}_{3}
\cr\end{matrix}\right)^{}_{\hspace{-0.08cm} \rm L} \hspace{0.05cm} \right]
W^-_\mu + {\rm h.c.} \; ,
\label{7}
\end{eqnarray}
where $U \equiv A U^{}_0$ is just the PMNS matrix in the seesaw framework because
the flavor eigenstates of three charged leptons have been taken to be
identical to their mass eigenstates from the very beginning. It is now
transparent that $U$ is not exactly unitary, and its deviation from $U^{}_0$
is measured by nonzero $R$ through $U U^\dagger = A A^\dagger = I - R R^\dagger$.

Substituting Eq.~(\ref{5}) into Eq.~(\ref{4}) allows us to arrive at the exact
seesaw relation between the light and heavy Majorana neutrino masses
\begin{eqnarray}
U D^{}_\nu U^T + R D^{}_N R^T = {\bf 0} \; ,
\label{8}
\end{eqnarray}
together with the expressions
\begin{eqnarray}
M^{}_{\rm D} \hspace{-0.2cm} & = & \hspace{-0.2cm}
R D^{}_N \left[ I - \left(B^{-1} S A^{-1} R\right)^T \right] U^{\prime T} \; ,
\nonumber \\
D^{}_{\rm R} \hspace{-0.2cm} & = & \hspace{-0.2cm}
U^\prime \left[ D^{}_N - \left(B^{-1} S A^{-1} R\right)
D^{}_N \left(B^{-1} S A^{-1} R\right)^T\right] U^{\prime T} \; ,
\hspace{0.6cm}
\label{9}
\end{eqnarray}
where $U^\prime \equiv U^\prime_0 B$ has been defined~\cite{Xing:2023adc}.
Given the fact that the elements of $U$ are at most of ${\cal O}(1)$ in
magnitude, Eq.~(\ref{8}) implies that the smallness of $m^{}_i/M^{}_i$
requires the smallness of the elements of $R$ in magnitude. So both $A$ and
$B$ are expected to be very close to the identity matrix $I$, as indicated by
Eq.~(\ref{6}). In this case the elements of $S$ must be of the same order as
those of $R$ in magnitude. Note
also that $U^\prime \to I$ can naturally be anticipated from Eq.~(\ref{9}),
simply because both $D^{}_{\rm R}$ and $D^{}_N$ are diagonal and real. We are
therefore left with $M^{}_{\rm D} \simeq R D^{}_N$ and $D^{}_{\rm R} \simeq D^{}_N$
as two excellent approximations.

Now that $Y^{}_\nu = M^{}_{\rm D}/\langle H\rangle \propto
R D^{}_N/\langle H\rangle$ holds, let us propose a generalized
CI parametrization of the Yukawa coupling matrix $Y^{}_\nu$ with
the help of the exact seesaw formula obtained in Eq.~(\ref{8}). The first
step is to express $R$ as follows:
\begin{eqnarray}
R = {\rm i} \hspace{0.05cm} U \sqrt{D^{}_\nu} \hspace{0.05cm}
\mathbb{O} \hspace{0.05cm} \frac{1}{\sqrt{D^{}_N}} \; ,
\label{10}
\end{eqnarray}
where $\mathbb{O}$ is an arbitrary complex orthogonal matrix. The second step is
to substitute Eq.~(\ref{10}) into Eq.~(\ref{9}), leading us to the Yukawa
coupling matrix
\begin{eqnarray}
Y^{}_\nu = \frac{\rm i}{\langle H\rangle} U \sqrt{D^{}_\nu} \hspace{0.05cm}
\mathbb{O} \hspace{0.05cm} \sqrt{D^{}_N} \hspace{0.05cm} I^\prime \; ,
\label{11}
\end{eqnarray}
where $I^\prime \equiv \left[ I - \left(B^{-1} S A^{-1} R\right)^T \right] U^{\prime T}$
is a dimensionless matrix which can approximate to $I$ in most cases. Comparing this
generalized CI parametrization with the original one shown in Eq.~(\ref{1}),
one may clearly see that two kinds of next-to-leading order effects (i.e., the deviation
of $U = A U^{}_0$ from $U^{}_0$ and that of $I^\prime$ from $I$) have been taken into
account.

Note that the unflavored thermal leptogenesis is essentially governed by the
imaginary part of $\left(Y^\dagger_\nu Y^{}_\nu\right)^2$, which appears in the
flavor-independent CP-violating asymmetries of the heavy Majorana neutrino
decays~\cite{Luty:1992un,Covi:1996wh,Plumacher:1996kc,Pilaftsis:1997jf}. Given the
above parametrization of $Y^{}_\nu$, we obtain
\footnote{Note that one usually calculates the decays of heavy Majorana neutrinos in
their mass basis taken in Eq.~(\ref{2}) before spontaneous gauge symmetry
breaking~\cite{Luty:1992un,Covi:1996wh,Plumacher:1996kc,Pilaftsis:1997jf}, instead
of in the seesaw mass basis for both light and heavy Majorana neutrinos obtained
from the transformation in Eq.~(\ref{4}). Namely, there is a small mismatch between
$D^{}_{\rm R}$ in Eq.~(\ref{2}) and $D^{}_N$ in Eq.~(\ref{4})~\cite{Xing:2023adc,
Drewes:2013gca,Canetti:2012kh,Drewes:2019byd}. For this reason, it is suitable to
omit the contributions of $I^\prime$, $U^{}_0$ and $S$ to $Y^\dagger_\nu Y^{}_\nu$
when discussing the issues of thermal leptogenesis in the seesaw mass basis.}
\begin{eqnarray}
Y^\dagger_\nu Y^{}_\nu \hspace{-0.2cm} & = & \hspace{-0.2cm}
\frac{1}{\langle H\rangle^2} I^{\prime \hspace{0.03cm} \dagger}
\sqrt{D^{}_N} \hspace{0.05cm} \mathbb{O}^\dagger \sqrt{D^{}_\nu} \left(I - U^\dagger_0
S^\dagger S U^{}_0\right) \sqrt{D^{}_\nu} \hspace{0.05cm} \mathbb{O} \sqrt{D^{}_N}
\hspace{0.05cm} I^\prime
\nonumber \\
\hspace{-0.2cm} & \simeq & \hspace{-0.2cm}
\frac{1}{\langle H\rangle^2}
\sqrt{D^{}_N} \hspace{0.05cm} \mathbb{O}^\dagger D^{}_\nu \mathbb{O} \sqrt{D^{}_N}
\label{12}
\end{eqnarray}
in the leading-order approximation. That is why some authors have concluded
that the CP-violating effects associated with the unflavored leptogenesis
scenario should solely be determined by the complex orthogonal matrix $\mathbb{O}$ and
have nothing to do with leptonic CP violation in neutrino oscillations characterized
by the Jarlskog invariant ${\cal J}^{}_\nu$ of the PMNS lepton flavor mixing
matrix $U \simeq U^{}_0$~\cite{Jarlskog:1985ht,Wu:1985ea,Cheng:1986in} (see,
e.g., Refs.~\cite{Ibarra:2003up,Pascoli:2006ie,Blanchet:2006be,Molinaro:2009lud,
Xing:2012gd,Bambhaniya:2016rbb,Xing:2020erm,Xing:2020ghj,Zhao:2021dwc,
Granelli:2023tcj,Zhao:2024uid,Zhao:2024mrs,Shao:2024fem,Xing:2009vb,
Rodejohann:2009cq,Antusch:2009gn} and references therein).

Such a conclusion is certainly problematic, as a correlation between the
unflavored leptogenesis and ${\cal J}^{}_\nu$ does not depend on whether the PMNS
matrix has appeared in the expression of $Y^\dagger_\nu Y^{}_\nu$ or not.
The point is that the CP-violating and lepton-number-violating decays of three
heavy Majorana neutrinos are governed by the {\it original} flavor parameters of
the seesaw mechanism, while those flavor parameters hidden in $D^{}_\nu$ and $U^{}_0$
are {\it derivational} in the sense that they must be determined by the same
{\it original} flavor parameters hidden in $A$, $R$ and $D^{}_N$ via the exact seesaw
relation $U^{}_0 D^{}_\nu U^T_0 = - \left(A^{-1} R\right) D^{}_N \left(A^{-1} R\right)^T$
as indicated by Eq.~(\ref{8})~\cite{Xing:2023adc,Xing:2023kdj,Xing:2024xwb}.

In fact, the complex orthogonal matrix $\mathbb{O}$ consists of both the original flavor
parameters hidden in $A$, $R$ and $D^{}_N$ and the derivational flavor parameters
hidden in $U^{}_0$ and $D^{}_\nu$:
\begin{eqnarray}
\mathbb{O} = -{\rm i} \left(U^{}_0 \sqrt{D^{}_\nu}\right)^{-1} \left(A^{-1}
R \sqrt{D^{}_N} \right) \; .
\label{13}
\end{eqnarray}
So taking a specific parameter space of $\mathbb{O}$ in calculating $Y^\dagger_\nu Y^{}_\nu$
for the thermal leptogenesis mechanism actually means a special assumption of the
correlations between the derivational and original flavor parameters in the canonical
seesaw framework.

\section{The complex orthogonal matrix $\mathbb{O}$}

To elucidate the above point of view more clearly, let us introduce a full Euler-like
parametrization of the flavor structure of the seesaw mechanism along the line of
thought illustrated in Eqs.~(\ref{4}) and (\ref{5}). We write out the three parts
of $\mathbb{U}$ in Eq.~(\ref{5}) as~\cite{Xing:2011ur}
\begin{eqnarray}
\left( \begin{matrix} U^{}_0 & 0 \cr 0 & I \cr \end{matrix} \right)
\hspace{-0.2cm} & = & \hspace{-0.2cm} O^{}_{23} O^{}_{13} O^{}_{12} \; ,
\nonumber \\
\left( \begin{matrix} I & 0 \cr 0 & U^{\prime}_0 \cr \end{matrix} \right)
\hspace{-0.2cm} & = & \hspace{-0.2cm} O^{}_{56} O^{}_{46} O^{}_{45} \; ,
\nonumber \\
\left( \begin{matrix} A & R \cr S & B \cr \end{matrix} \right)
\hspace{-0.2cm} & = & \hspace{-0.2cm} O^{}_{36} O^{}_{26} O^{}_{16}
O^{}_{35} O^{}_{25} O^{}_{15} O^{}_{34} O^{}_{24} O^{}_{14} \; , \hspace{0.7cm}
\label{14}
\end{eqnarray}
where each of the fifteen $6 \times 6$ unitary matrices $O^{}_{ij}$
(for $1 \leq i < j \leq 6$) has the following properties: its $(i, i)$ and
$(j, j)$ entries are both identical to $c^{}_{ij} \equiv \cos\theta^{}_{ij}$
with $\theta^{}_{ij}$ being a flavor mixing angle and lying in the first quadrant,
its other four diagonal elements are all equal to one, its $(i, j)$ and $(j, i)$
entries are respectively set as $\hat{s}^{*}_{ij} \equiv e^{-{\rm i}\delta^{}_{ij}}
\sin\theta^{}_{ij}$ and $-\hat{s}^{}_{ij} \equiv -e^{{\rm i}\delta^{}_{ij}}
\sin\theta^{}_{ij}$ with $\delta^{}_{ij}$ being a CP-violating phase, and its
other off-diagonal elements are all vanishing. Then the explicit expressions of
$A$, $R$ and $U^{}_0$ are found to be
\begin{eqnarray}
A \hspace{-0.2cm} & = & \hspace{-0.2cm}
\left( \begin{matrix} c^{}_{14} c^{}_{15} c^{}_{16} & 0 & 0
\cr \vspace{-0.45cm} \cr
\begin{array}{l} -c^{}_{14} c^{}_{15} \hat{s}^{}_{16} \hat{s}^*_{26} -
c^{}_{14} \hat{s}^{}_{15} \hat{s}^*_{25} c^{}_{26} \\
-\hat{s}^{}_{14} \hat{s}^*_{24} c^{}_{25} c^{}_{26} \end{array} &
c^{}_{24} c^{}_{25} c^{}_{26} & 0 \cr \vspace{-0.45cm} \cr
\begin{array}{l} -c^{}_{14} c^{}_{15} \hat{s}^{}_{16} c^{}_{26} \hat{s}^*_{36}
+ c^{}_{14} \hat{s}^{}_{15} \hat{s}^*_{25} \hat{s}^{}_{26} \hat{s}^*_{36} \\
- c^{}_{14} \hat{s}^{}_{15} c^{}_{25} \hat{s}^*_{35} c^{}_{36} +
\hat{s}^{}_{14} \hat{s}^*_{24} c^{}_{25} \hat{s}^{}_{26}
\hat{s}^*_{36} \\
+ \hat{s}^{}_{14} \hat{s}^*_{24} \hat{s}^{}_{25} \hat{s}^*_{35}
c^{}_{36} - \hat{s}^{}_{14} c^{}_{24} \hat{s}^*_{34} c^{}_{35}
c^{}_{36} \end{array} &
\begin{array}{l} -c^{}_{24} c^{}_{25} \hat{s}^{}_{26} \hat{s}^*_{36} -
c^{}_{24} \hat{s}^{}_{25} \hat{s}^*_{35} c^{}_{36} \\
-\hat{s}^{}_{24} \hat{s}^*_{34} c^{}_{35} c^{}_{36} \end{array} &
c^{}_{34} c^{}_{35} c^{}_{36} \cr \end{matrix} \right) \; ,
\nonumber \\
R \hspace{-0.2cm} & = & \hspace{-0.2cm}
\left( \begin{matrix} \hat{s}^*_{14} c^{}_{15} c^{}_{16} &
\hat{s}^*_{15} c^{}_{16} & \hat{s}^*_{16} \cr \vspace{-0.45cm} \cr
\begin{array}{l} -\hat{s}^*_{14} c^{}_{15} \hat{s}^{}_{16} \hat{s}^*_{26} -
\hat{s}^*_{14} \hat{s}^{}_{15} \hat{s}^*_{25} c^{}_{26} \\
+ c^{}_{14} \hat{s}^*_{24} c^{}_{25} c^{}_{26} \end{array} & -
\hat{s}^*_{15} \hat{s}^{}_{16} \hat{s}^*_{26} + c^{}_{15}
\hat{s}^*_{25} c^{}_{26} & c^{}_{16} \hat{s}^*_{26} \cr \vspace{-0.45cm} \cr
\begin{array}{l} -\hat{s}^*_{14} c^{}_{15} \hat{s}^{}_{16} c^{}_{26}
\hat{s}^*_{36} + \hat{s}^*_{14} \hat{s}^{}_{15} \hat{s}^*_{25}
\hat{s}^{}_{26} \hat{s}^*_{36} \\ - \hat{s}^*_{14} \hat{s}^{}_{15}
c^{}_{25} \hat{s}^*_{35} c^{}_{36} - c^{}_{14} \hat{s}^*_{24}
c^{}_{25} \hat{s}^{}_{26}
\hat{s}^*_{36} \\
- c^{}_{14} \hat{s}^*_{24} \hat{s}^{}_{25} \hat{s}^*_{35}
c^{}_{36} + c^{}_{14} c^{}_{24} \hat{s}^*_{34} c^{}_{35} c^{}_{36}
\end{array} &
\begin{array}{l} -\hat{s}^*_{15} \hat{s}^{}_{16} c^{}_{26} \hat{s}^*_{36}
- c^{}_{15} \hat{s}^*_{25} \hat{s}^{}_{26} \hat{s}^*_{36} \\
+c^{}_{15} c^{}_{25} \hat{s}^*_{35} c^{}_{36} \end{array} &
c^{}_{16} c^{}_{26} \hat{s}^*_{36} \cr \end{matrix} \right) \; , \hspace{0.6cm}
\label{15}
\end{eqnarray}
and
\begin{eqnarray}
U^{}_0 \hspace{-0.2cm} & = & \hspace{-0.2cm}
\left( \begin{matrix} c^{}_{12} c^{}_{13} & \hat{s}^*_{12}
c^{}_{13} & \hat{s}^*_{13} \cr
-\hat{s}^{}_{12} c^{}_{23} -
c^{}_{12} \hat{s}^{}_{13} \hat{s}^*_{23} & c^{}_{12} c^{}_{23} -
\hat{s}^*_{12} \hat{s}^{}_{13} \hat{s}^*_{23} & c^{}_{13}
\hat{s}^*_{23} \cr
\hat{s}^{}_{12} \hat{s}^{}_{23} - c^{}_{12}
\hat{s}^{}_{13} c^{}_{23} & -c^{}_{12} \hat{s}^{}_{23} -
\hat{s}^*_{12} \hat{s}^{}_{13} c^{}_{23} & c^{}_{13} c^{}_{23}
\cr \end{matrix} \right) \; . \hspace{0.6cm}
\label{16}
\end{eqnarray}
The expressions of $B$ and $S$ can be respectively read off from those of $A^*$
and $-R^*$ with the subscript replacements $15 \leftrightarrow 24$,
$16 \leftrightarrow 34$ and $26 \leftrightarrow 35$; and the expression of
$U^\prime_0$ can similarly be obtained from that of $U^{}_0$
with the subscript replacements $12 \leftrightarrow 45$, $13 \leftrightarrow 46$
and $23 \leftrightarrow 56$. As both $U = A U^{}_0$ and $R$ appear in the
leptonic weak charged-current interactions given in Eq.~(\ref{7}),
one may always redefine the phases of three charged-lepton fields to remove
three of the nine phase parameters (or their combinations) of $A$ and $R$
such that only six independent CP-violating phases are
left~\cite{Endoh:2000hc,Branco:2001pq}.

Given the fact that the Yukawa coupling matrix $Y^{}_\nu$ is dominated by
$R$ and $D^{}_N$, the original seesaw flavor parameters can therefore be
identified as the eighteen independent parameters of $R$ and $D^{}_N$: three
heavy Majorana neutrino masses ($M^{}_i$ for $i = 1, 2, 3$), nine active-sterile
flavor mixing angles ($\theta^{}_{ij}$ for $i = 1, 2, 3$ and $j = 4, 5, 6$)
and six independent CP-violating phases ($\delta^{}_{ij}$ for $i = 1, 2, 3$
and $j = 4, 5, 6$). In comparison, the flavor parameters hidden in $U^\prime_0$
are sterile in the sense that they do not contribute to any available physical
processes, and the flavor parameters of $U^{}_0$ and $D^{}_\nu$ for three active
Majorana neutrinos can in principle be determined from those of $R$ and $D^{}_N$
via the exact seesaw formula. So the complex orthogonal matrix $\mathbb{O}$ in
Eq.~(\ref{13}) is actually a nonlinear function of all the original seesaw flavor
parameters.

Substituting Eqs.~(\ref{14}), (\ref{15}) and (\ref{16}) into Eq.~(\ref{13}), one
may derive the exact expressions of $\mathbb{O}$ in terms of the original and
derivational flavor parameters of the canonical seesaw mechanism
\footnote{It will certainly be much better if all the derivational parameters
in $D^{}_\nu$ and $U^{}_0$ can be expressed in terms of those original
parameters in $D^{}_N$, $A$ and $R$. But such expressions are too complicated
to be available at present.}.
We find that the elements of $\mathbb{O}$ can be expressed as
$\mathbb{O}^{}_{ij} = \sqrt{M^{}_j/m^{}_i} \hspace{0.05cm} F^{}_{ij}$ with
\begin{eqnarray}
F^{}_{ij} = -{\rm i} \hspace{0.05cm} \sum^3_{k = 1} \left[\left(U^*_0\right)^{}_{k i}
\left(A^{-1} R\right)^{}_{k j}\right] \; ,
\label{17}
\end{eqnarray}
where $i$, $j$ and $k$ run over $1$, $2$ and $3$. The exact results of
$F^{}_{ij}$ turn out to be too lengthy to be useful. As the active-sterile
flavor mixing angles in $A$ and $R$ measure the strength of non-unitarity of the
PMNS matrix $U$ (i.e., $UU^\dagger = A A^\dagger = I - RR^\dagger \neq I$), they
must be small enough. The latest in-depth analysis of currently available
electroweak precision measurements and neutrino oscillation data has set a
rather strong bound on possible non-unitarity of $U$ --- the latter should be
below ${\cal O}(10^{-3})$ in the canonical seesaw framework~\cite{Blennow:2023mqx},
and hence each of the nine active-sterile flavor mixing angles of $R$ is
expected to be smaller than ${\cal O}(10^{-1.5})$. So we may make a
very good approximation
\begin{eqnarray}
A^{-1} R =
\left(\begin{matrix} \hat{s}^*_{14} & \hat{s}^*_{15} & \hat{s}^*_{16} \cr
\hat{s}^*_{24} & \hat{s}^*_{25} & \hat{s}^*_{26} \cr
\hat{s}^*_{34} & \hat{s}^*_{35} & \hat{s}^*_{36} \cr \end{matrix}\right)
+ {\cal O} \left(s^3_{ij}\right)
\label{18}
\end{eqnarray}
in our subsequent calculations (for $i = 1, 2, 3$ and $j = 4, 5, 6$).
Explicitly, we have
\begin{align}
{\rm i} \hspace{0.05cm} F^{}_{11} \simeq & \hspace{0.15cm}
c^{}_{12} c^{}_{13} \hat{s}^*_{14} - \left(\hat{s}^*_{12} c^{}_{23} +
c^{}_{12} \hat{s}^*_{13} \hat{s}^{}_{23}\right) \hat{s}^*_{24} +
\left(\hat{s}^*_{12} \hat{s}^{*}_{23} -
c^{}_{12} \hat{s}^*_{13} c^{}_{23}\right) \hat{s}^*_{34} \; ,
\nonumber \\
{\rm i} \hspace{0.05cm} F^{}_{21} \simeq & \hspace{0.15cm}
\hat{s}^{}_{12} c^{}_{13} \hat{s}^*_{14} + \left(c^{}_{12} c^{}_{23} -
\hat{s}^{}_{12} \hat{s}^*_{13} \hat{s}^{}_{23}\right) \hat{s}^*_{24} -
\left(c^{}_{12} \hat{s}^{*}_{23} +
\hat{s}^{}_{12} \hat{s}^*_{13} c^{}_{23}\right) \hat{s}^*_{34} \; ,
\nonumber \\
{\rm i} \hspace{0.05cm} F^{}_{31} \simeq & \hspace{0.15cm}
\hat{s}^{}_{13} \hat{s}^*_{14} + c^{}_{13} \hat{s}^{}_{23} \hat{s}^*_{24} +
c^{}_{13} c^{}_{23} \hat{s}^*_{34} \; ;
\nonumber \\
{\rm i} \hspace{0.05cm} F^{}_{12} \simeq & \hspace{0.15cm}
c^{}_{12} c^{}_{13} \hat{s}^*_{15} - \left(\hat{s}^*_{12} c^{}_{23} +
c^{}_{12} \hat{s}^*_{13} \hat{s}^{}_{23}\right) \hat{s}^*_{25} +
\left(\hat{s}^*_{12} \hat{s}^{*}_{23} -
c^{}_{12} \hat{s}^*_{13} c^{}_{23}\right) \hat{s}^*_{35} \; ,
\nonumber \\
{\rm i} \hspace{0.05cm} F^{}_{22} \simeq & \hspace{0.15cm}
\hat{s}^{}_{12} c^{}_{13} \hat{s}^*_{15} + \left(c^{}_{12} c^{}_{23} -
\hat{s}^{}_{12} \hat{s}^*_{13} \hat{s}^{}_{23}\right) \hat{s}^*_{25} -
\left(c^{}_{12} \hat{s}^{*}_{23} +
\hat{s}^{}_{12} \hat{s}^*_{13} c^{}_{23}\right) \hat{s}^*_{35} \; ,
\nonumber \\
{\rm i} \hspace{0.05cm} F^{}_{32} \simeq & \hspace{0.15cm}
\hat{s}^{}_{13} \hat{s}^*_{15} + c^{}_{13} \hat{s}^{}_{23} \hat{s}^*_{25} +
c^{}_{13} c^{}_{23} \hat{s}^*_{35} \; ;
\nonumber \\
{\rm i} \hspace{0.05cm} F^{}_{13} \simeq & \hspace{0.15cm}
c^{}_{12} c^{}_{13} \hat{s}^*_{16} - \left(\hat{s}^*_{12} c^{}_{23} +
c^{}_{12} \hat{s}^*_{13} \hat{s}^{}_{23}\right) \hat{s}^*_{26} +
\left(\hat{s}^*_{12} \hat{s}^{*}_{23} -
c^{}_{12} \hat{s}^*_{13} c^{}_{23}\right) \hat{s}^*_{36} \; ,
\nonumber \\
{\rm i} \hspace{0.05cm} F^{}_{23} \simeq & \hspace{0.15cm}
\hat{s}^{}_{12} c^{}_{13} \hat{s}^*_{16} + \left(c^{}_{12} c^{}_{23} -
\hat{s}^{}_{12} \hat{s}^*_{13} \hat{s}^{}_{23}\right) \hat{s}^*_{26} -
\left(c^{}_{12} \hat{s}^{*}_{23} +
\hat{s}^{}_{12} \hat{s}^*_{13} c^{}_{23}\right) \hat{s}^*_{36} \; , \hspace{0.8cm}
\nonumber \\
{\rm i} \hspace{0.05cm} F^{}_{33} \simeq & \hspace{0.15cm}
\hat{s}^{}_{13} \hat{s}^*_{16} + c^{}_{13} \hat{s}^{}_{23} \hat{s}^*_{26} +
c^{}_{13} c^{}_{23} \hat{s}^*_{36} \; .
\label{19}
\end{align}
The elements $\mathbb{O}^{}_{ij}$ can then be determined from $F^{}_{ij}$ multiplied
by the factor $\sqrt{M^{}_j/m^{}_i}$ (for $i, j = 1, 2, 3$).

Note that Eq.~(\ref{19}) allows us to express the nine active-sterile flavor
mixing parameters as follows:
\begin{align}
\hat{s}^*_{14} \simeq & \hspace{0.15cm} {\rm i} \hspace{0.05cm}
\sqrt{\frac{m^{}_1}{M^{}_1}} \left[ c^{}_{12} c^{}_{13} \mathbb{O}^{}_{11} +
\sqrt{\frac{m^{}_2}{m^{}_1}} \hspace{0.05cm} \hat{s}^*_{12} c^{}_{13}
\mathbb{O}^{}_{21} + \sqrt{\frac{m^{}_3}{m^{}_1}} \hspace{0.05cm} \hat{s}^*_{13}
\mathbb{O}^{}_{31} \right] \; ,
\nonumber \\
\hat{s}^*_{24} \simeq & \hspace{0.15cm} {\rm i} \hspace{0.05cm}
\sqrt{\frac{m^{}_1}{M^{}_1}} \left[ -\left(\hat{s}^{}_{12} c^{}_{23}
+ c^{}_{12} \hat{s}^{}_{13} \hat{s}^*_{23}\right) \mathbb{O}^{}_{11} +
\sqrt{\frac{m^{}_2}{m^{}_1}} \left(c^{}_{12} c^{}_{23} - \hat{s}^*_{12}
\hat{s}^{}_{13} \hat{s}^*_{23}\right) \mathbb{O}^{}_{21} + \sqrt{\frac{m^{}_3}{m^{}_1}}
\hspace{0.05cm} c^{}_{13} \hat{s}^*_{23} \mathbb{O}^{}_{31} \right] \; ,
\nonumber \\
\hat{s}^*_{34} \simeq & \hspace{0.15cm} {\rm i} \hspace{0.05cm}
\sqrt{\frac{m^{}_1}{M^{}_1}} \left[ \left(\hat{s}^{}_{12} \hat{s}^{}_{23}
- c^{}_{12} \hat{s}^{}_{13} c^{}_{23}\right) \mathbb{O}^{}_{11} -
\sqrt{\frac{m^{}_2}{m^{}_1}} \left(c^{}_{12} \hat{s}^{}_{23} + \hat{s}^*_{12}
\hat{s}^{}_{13} c^{}_{23}\right) \mathbb{O}^{}_{21} + \sqrt{\frac{m^{}_3}{m^{}_1}}
\hspace{0.05cm} c^{}_{13} c^{}_{23} \mathbb{O}^{}_{31} \right] \; ;
\nonumber \\
\hat{s}^*_{15} \simeq & \hspace{0.15cm} {\rm i} \hspace{0.05cm}
\sqrt{\frac{m^{}_1}{M^{}_2}} \left[ c^{}_{12} c^{}_{13} \mathbb{O}^{}_{12} +
\sqrt{\frac{m^{}_2}{m^{}_1}} \hspace{0.05cm} \hat{s}^*_{12} c^{}_{13}
\mathbb{O}^{}_{22} + \sqrt{\frac{m^{}_3}{m^{}_1}} \hspace{0.05cm} \hat{s}^*_{13}
\mathbb{O}^{}_{32} \right] \; ,
\nonumber \\
\hat{s}^*_{25} \simeq & \hspace{0.15cm} {\rm i} \hspace{0.05cm}
\sqrt{\frac{m^{}_1}{M^{}_2}} \left[ -\left(\hat{s}^{}_{12} c^{}_{23}
+ c^{}_{12} \hat{s}^{}_{13} \hat{s}^*_{23}\right) \mathbb{O}^{}_{12} +
\sqrt{\frac{m^{}_2}{m^{}_1}} \left(c^{}_{12} c^{}_{23} - \hat{s}^*_{12}
\hat{s}^{}_{13} \hat{s}^*_{23}\right) \mathbb{O}^{}_{22} + \sqrt{\frac{m^{}_3}{m^{}_1}}
\hspace{0.05cm} c^{}_{13} \hat{s}^*_{23} \mathbb{O}^{}_{32} \right] \; ,
\nonumber \\
\hat{s}^*_{35} \simeq & \hspace{0.15cm} {\rm i} \hspace{0.05cm}
\sqrt{\frac{m^{}_1}{M^{}_2}} \left[ \left(\hat{s}^{}_{12} \hat{s}^{}_{23}
- c^{}_{12} \hat{s}^{}_{13} c^{}_{23}\right) \mathbb{O}^{}_{12} -
\sqrt{\frac{m^{}_2}{m^{}_1}} \left(c^{}_{12} \hat{s}^{}_{23} + \hat{s}^*_{12}
\hat{s}^{}_{13} c^{}_{23}\right) \mathbb{O}^{}_{22} + \sqrt{\frac{m^{}_3}{m^{}_1}}
\hspace{0.05cm} c^{}_{13} c^{}_{23} \mathbb{O}^{}_{32} \right] \; ;
\nonumber \\
\hat{s}^*_{16} \simeq & \hspace{0.15cm} {\rm i} \hspace{0.05cm}
\sqrt{\frac{m^{}_1}{M^{}_3}} \left[ c^{}_{12} c^{}_{13} \mathbb{O}^{}_{13} +
\sqrt{\frac{m^{}_2}{m^{}_1}} \hspace{0.05cm} \hat{s}^*_{12} c^{}_{13}
\mathbb{O}^{}_{23} + \sqrt{\frac{m^{}_3}{m^{}_1}} \hspace{0.05cm} \hat{s}^*_{13}
\mathbb{O}^{}_{33} \right] \; ,
\nonumber \\
\hat{s}^*_{26} \simeq & \hspace{0.15cm} {\rm i} \hspace{0.05cm}
\sqrt{\frac{m^{}_1}{M^{}_3}} \left[ -\left(\hat{s}^{}_{12} c^{}_{23}
+ c^{}_{12} \hat{s}^{}_{13} \hat{s}^*_{23}\right) \mathbb{O}^{}_{13} +
\sqrt{\frac{m^{}_2}{m^{}_1}} \left(c^{}_{12} c^{}_{23} - \hat{s}^*_{12}
\hat{s}^{}_{13} \hat{s}^*_{23}\right) \mathbb{O}^{}_{23} + \sqrt{\frac{m^{}_3}{m^{}_1}}
\hspace{0.05cm} c^{}_{13} \hat{s}^*_{23} \mathbb{O}^{}_{33} \right] \; ,
\nonumber \\
\hat{s}^*_{36} \simeq & \hspace{0.15cm} {\rm i} \hspace{0.05cm}
\sqrt{\frac{m^{}_1}{M^{}_3}} \left[ \left(\hat{s}^{}_{12} \hat{s}^{}_{23}
- c^{}_{12} \hat{s}^{}_{13} c^{}_{23}\right) \mathbb{O}^{}_{13} -
\sqrt{\frac{m^{}_2}{m^{}_1}} \left(c^{}_{12} \hat{s}^{}_{23} + \hat{s}^*_{12}
\hat{s}^{}_{13} c^{}_{23}\right) \mathbb{O}^{}_{23} + \sqrt{\frac{m^{}_3}{m^{}_1}}
\hspace{0.05cm} c^{}_{13} c^{}_{23} \mathbb{O}^{}_{33} \right] \; .
\label{20}
\end{align}
So assuming a specific pattern of $\mathbb{O}$ will definitely lead to very strong
constraints on the correlations between the original and derivational flavor
parameters of the seesaw mechanism. For example, the assumption of $\mathbb{O} = I$
will make Eq.~(\ref{20}) greatly simplified:
\begin{align}
\hat{s}^*_{14} \simeq & \hspace{0.15cm} {\rm i} \hspace{0.05cm}
\sqrt{\frac{m^{}_1}{M^{}_1}} \hspace{0.05cm} c^{}_{12} c^{}_{13} \; ,
\quad
\hat{s}^*_{24} \simeq -{\rm i} \hspace{0.05cm}
\sqrt{\frac{m^{}_1}{M^{}_1}} \left(\hat{s}^{}_{12} c^{}_{23}
+ c^{}_{12} \hat{s}^{}_{13} \hat{s}^*_{23}\right) \; ,
\quad
\hat{s}^*_{34} \simeq {\rm i} \hspace{0.05cm}
\sqrt{\frac{m^{}_1}{M^{}_1}} \left(\hat{s}^{}_{12} \hat{s}^{}_{23}
- c^{}_{12} \hat{s}^{}_{13} c^{}_{23}\right) \; ;
\nonumber \\
\hat{s}^*_{15} \simeq & \hspace{0.15cm} {\rm i} \hspace{0.05cm}
\sqrt{\frac{m^{}_2}{M^{}_2}} \hspace{0.05cm} \hat{s}^*_{12} c^{}_{13} \; ,
\quad
\hat{s}^*_{25} \simeq {\rm i} \hspace{0.05cm}
\sqrt{\frac{m^{}_2}{M^{}_2}} \left(c^{}_{12} c^{}_{23} - \hat{s}^*_{12}
\hat{s}^{}_{13} \hat{s}^*_{23}\right) \; ,
\quad
\hat{s}^*_{35} \simeq -{\rm i} \hspace{0.05cm}
\sqrt{\frac{m^{}_2}{M^{}_2}} \left(c^{}_{12} \hat{s}^{}_{23} + \hat{s}^*_{12}
\hat{s}^{}_{13} c^{}_{23}\right) \; ;
\nonumber \\
\hat{s}^*_{16} \simeq & \hspace{0.15cm} {\rm i} \hspace{0.05cm}
\sqrt{\frac{m^{}_3}{M^{}_3}} \hspace{0.05cm} \hat{s}^*_{13} \; ,
\quad
\hat{s}^*_{26} \simeq {\rm i} \hspace{0.05cm}
\sqrt{\frac{m^{}_3}{M^{}_3}} \hspace{0.05cm} c^{}_{13} \hat{s}^*_{23} \; ,
\quad
\hat{s}^*_{36} \simeq {\rm i} \hspace{0.05cm}
\sqrt{\frac{m^{}_3}{M^{}_3}} \hspace{0.05cm} c^{}_{13} c^{}_{23} \; ,
\label{21}
\end{align}
which can also be obtained from a simplification of Eq.~(\ref{13}) in
this case (i.e., $R = {\rm i} A U^{}_0 \sqrt{D^{}_\nu/D^{}_N}$). No
matter what form of the complex orthogonal matrix $\mathbb{O}$ is conjectured,
however, one should always keep in mind that the flavor parameters $m^{}_i$,
$\theta^{}_{ij}$ and $\delta^{}_{ij}$ (for $i, j = 1, 2, 3$) for the three light
Majorana neutrinos are {\it derivational} in the sense that they are absent in
the seesaw Lagrangian in Eq.~(\ref{2}) and originate from the seesaw relation
in Eq.~(\ref{8}).

\section{Summary and remarks}

We have generalized the CI parametrization for the Yukawa interactions of
Majorana neutrino by starting from the exact seesaw relation and taking
account of possible non-unitarity of the $3 \times 3$ PMNS lepton flavor
mixing matrix. The complex orthogonal matrix $\mathbb{O}$ in this
parametrization has been derived with the help of a full $6 \times 6$
Euler-like block description of the flavor structure for the canonical
seesaw mechanism. We find $\mathbb{O}^{}_{ij} = \sqrt{M^{}_j/m^{}_i}
\hspace{0.05cm} F^{}_{ij}$ with $m^{}_i$ and $M^{}_j$ being the masses of
light and heavy Majorana neutrinos and $F^{}_{ij}$ consisting of the PMNS
and active-sterile flavor mixing parameters (for $i, j = 1, 2, 3$). This
result means that assuming a specific pattern of $\mathbb{O}$ is equivalent
to imposing some special conditions on the seesaw parameter space.

Let us make some further comments on the physical meaning of $\mathbb{O}$
in the CI parametrization and its phenomenological applications.
\begin{itemize}
\item     One may certainly argue that a given pattern of $\mathbb{O}$
corresponds to a given texture of the Yukawa coupling matrix $Y^{}_\nu$,
and thus they should have the same physical implications. However,
$\mathbb{O}$ itself does not show up in the seesaw Lagrangian, making it
inconvenient to determine or constrain the texture of $\mathbb{O}$ with a
kind of flavor symmetry in a concrete neutrino mass model~\cite{Xing:2020ijf}.
That is why the CI parametrization has usually been used to explore the
whole parameter space of $Y^{}_\nu$ (or a part of it) by adjusting those
free parameters of $\mathbb{O}$ in the numerical analysis of a specific seesaw
or leptogenesis model.

\item     As we have shown in Eq.~(\ref{13}), the complex orthogonal
matrix $\mathbb{O}$ apparently consists of both the original and derivational
flavor parameters in the canonical seesaw framework. So assuming a specific
texture of $\mathbb{O}$ is equivalent to assuming some special parameter
correlations, as illustrated in section 3. Given that all the
original seesaw parameters hidden in $D^{}_N$ and $R$ are unknown, such
assumptions have no problem in the general case. But for a concrete seesaw
model with the structure of $Y^{}_\nu$ having partly been constrained, one
should take care of the issue of self-consistency when employing the CI
parametrization and allowing $\mathbb{O}$ to take random values.

\item     Note that allowing $\mathbb{O}$ to arbitrarily vary in a numerical
calculation of $Y^{}_\nu$ will definitely ``dilute" the contributions
of those input values of $m^{}_i$, $\theta^{}_{ij}$ and $\delta^{}_{ij}$ (for
$i, j = 1, 2, 3$) to $Y^{}_\nu$, as can be clearly seen in Eq.~(\ref{20}).
In other words, the used experimentally-allowed ranges of the masses and flavor
mixing parameters for the three light Majorana neutrinos would be partly
``weakened" when the nine elements of $\mathbb{O}$ are numerically adjusted,
making the claimed dependence of $Y^{}_\nu$ on the light degrees of freedom
too {\it quantitatively} sensitive to be instructive. In this situation it should 
be more convenient to adopt a direct parametrization of $Y^{}_\nu$ itself and 
connect its free parameters to $D^{}_\nu$ and $U$ via the exact seesaw formula 
in Eq.~(\ref{8}).
\end{itemize}
Of course, our analytical results obtained in section 3 can be simplified
in the minimal seesaw mechanism with only two right-handed neutrino 
fields (see Ref.~\cite{Xing:2020ald} for a recent review)
\footnote{It will also be interesting to extend the similar discussions to
a more general parametrization of the seesaw models with any number
of right-handed neutrino fields, as proposed in
Refs.~\cite{Bortolato:2020bgy,Cordero-Carrion:2018xre,Cordero-Carrion:2019qtu}.}.

Finally, we emphasize that the flavor-independent CP-violating asymmetries
between $N^{}_i \to \ell + H$ and $N^{}_i \to \overline{\ell} + \overline{H}$
decays as allowed by the Yukawa interactions in Eq.~(\ref{2}) depend only
on the original seesaw flavor parameters,
\begin{align}
\varepsilon^{}_{i} \equiv & \hspace{0.13cm}
\frac{\displaystyle \sum^{}_\alpha \Big[\Gamma({N}^{}_i \to \ell^{}_\alpha + H)
- \Gamma({N}^{}_i \to \overline{\ell^{}_\alpha} +
\overline{H})\Big]}{\displaystyle \sum_\alpha \Big[\Gamma({N}^{}_i \to
\ell^{}_\alpha + H) + \Gamma({N}^{}_i \to \overline{\ell^{}_\alpha}
+ \overline{H})\Big]}
\nonumber \\
\simeq & \hspace{0.13cm}
\frac{1}{\displaystyle 8\pi \langle H\rangle^2 \sum_\alpha \left|R^{}_{\alpha i}\right|^2}
\sum^3_{j = 1} \Bigg\{ M^2_j \hspace{0.1cm} {\rm Im} \Bigg[
\sum_\alpha \left(R^*_{\alpha i} R^{}_{\alpha j}\right)\Bigg]^2 \xi(x^{}_{ji}) \Bigg\} \; ,
\hspace{0.5cm}
\label{22}
\end{align}
where $\xi(x^{}_{ji}) = \sqrt{x^{}_{ji}} \left\{1 + 1 /
\left(1 - x^{}_{ji}\right) + \left(1 + x^{}_{ji}\right) \ln \left[x^{}_{ji}
/ \left(1 + x^{}_{ji}\right)\right] \right\}$ with $x^{}_{ji} \equiv
{M}^2_j/{M}^2_i$ are the loop functions, and the subscripts ``$i$' and ``$\alpha$"
run respectively over $(1, 2, 3)$ and $(e, \mu, \tau)$. There is certainly no
place for the flavor parameters of three light Majorana neutrinos in Eq.~(\ref{22}),
as the electroweak gauge symmetry is unbroken. However, there must be a clear
correlation between the CP-violating asymmetries $\varepsilon^{}_i$ and
the Jarlskog invariant ${\cal J}^{}_\nu$, once the latter is also calculated in
terms of $R$ via the seesaw relation~\cite{Xing:2023kdj,Xing:2024xwb}. In this
regard one may therefore avoid possible ambiguities associated with the absence of
$U^{}_0$ and the presence of $\mathbb{O}$ in the expression of $\varepsilon^{}_i$
when applying the CI parametrization to the issues of thermal leptogenesis.

\section*{Acknowledgements}

This research work is supported in part by the National Natural Science
Foundation of China under grant No. 12075254.


\end{document}